\documentclass[twocolumn,notitlepage,nofootinbib]{revtex4-2}
\usepackage{amsmath,amssymb,bm,graphicx,hyperref}
\allowdisplaybreaks[1]

\renewcommand\d{\partial}
\newcommand\grad{\bm{\nabla}}
\newcommand\+{\dagger}
\newcommand\<{\langle}
\renewcommand\>{\rangle}
\newcommand\eps{\varepsilon}
\newcommand\ep{\eps_p}
\newcommand\eq{\eps_q}
\newcommand\f{\bar{f}}
\newcommand\g{\bar{g}}
\renewcommand\r{\bm{r}}
\newcommand\E{\mathcal{E}}
\newcommand\G{\mathcal{G}}
\renewcommand\H{\mathcal{H}}
\newcommand\J{\mathcal{J}}
\newcommand\K{\mathcal{K}}
\newcommand\N{\mathcal{N}}
\renewcommand\P{\mathcal{P}}
\newcommand\Q{\mathcal{Q}}
\newcommand\T{\mathcal{T}}
\newcommand\W{\mathcal{W}}
\newcommand\R{\mathbb{R}}
\newcommand\Z{\mathbb{Z}}
\newcommand\cl{\mathrm{coll}}
\newcommand\hd{\mathrm{hydro}}
\newcommand\AL{\mathrm{AL}}
\newcommand\MT{\mathrm{MT}}
\newcommand\LHS{\mathrm{LHS}}
\newcommand\RHS{\mathrm{RHS}}
\DeclareMathOperator\Li{Li}

\let\Im\relax\DeclareMathOperator\Im{Im}

\begin{document}

\title{Thermal conductivity of a weakly interacting Bose gas in quasi-one-dimension}

\author{Tomohiro Tanaka}
\author{Yusuke Nishida}
\affiliation{Department of Physics, Tokyo Institute of Technology,
Ookayama, Meguro, Tokyo 152-8551, Japan}

\date{March 2022}

\begin{abstract}
Transport coefficients are typically divergent for quantum integrable systems in one dimension, such as a Bose gas with a two-body contact interaction.
However, when a one-dimensional system is realized by confining bosons into a tight matter waveguide, an effective three-body interaction inevitably arises as leading perturbation to break the integrability.
This fact motivates us to study the thermal conductivity of a Bose gas in one dimension with both two-body and three-body interactions.
In particular, we evaluate the Kubo formula exactly to the lowest order in perturbation by summing up all contributions that are naively higher orders in perturbation but become comparable in the zero-frequency limit due to the pinch singularity.
Consequently, a self-consistent equation for a vertex function is derived, showing that the thermal conductivity in quasi-one-dimension is dominated by the three-body interaction rather than the two-body interaction.
Furthermore, the resulting thermal conductivity in the weak-coupling limit proves to be identical to that computed based on the quantum Boltzmann equation and its temperature dependence is numerically determined.
\end{abstract}

\maketitle
\tableofcontents

\section{Introduction}
A nonvanishing Drude weight indicates a divergent transport coefficient at zero frequency and serves as diagnostics of whether the transport is ballistic or diffusive~\cite{Kohn:1964}.
As far as fluids with translational invariance are concerned, the Drude weight for mass transport is always nonvanishing because the mass current (i.e., momentum) is conserved~\cite{Mahan}, whereas that for energy transport typically vanishes if particles are interacting.
However, quantum integrable systems in one dimension have been found so exceptional that their Drude weights remain nonvanishing due to a macroscopic number of conservation laws~\cite{Castella:1995,Zotos:1997}.
One such system is a Bose gas with a two-body contact interaction known as the Lieb-Liniger model~\cite{Lieb:1963a,Lieb:1963b}, whose Drude weight was derived with the thermodynamic Bethe ansatz~\cite{Doyon:2017}.

An ideal platform for experimental study of the Lieb-Liniger model has been provided by ultracold atoms~\cite{Bloch:2008,Cazalilla:2011}, where anomalous nonequilibrium dynamics rooted in the integrability was observed~\cite{Kinoshita:2006}.
However, when a one-dimensional system is realized by confining bosons into a tight matter waveguide, effective multibody interactions inevitably arise from virtual transverse excitations in spite of their interaction being purely pairwise in free space~\cite{Muryshev:2002,Sinha:2006,Mazets:2008}.
Although such multibody interactions are usually neglected in a dilute gas, the three-body interaction may cause significant consequences for transport properties because it is the leading perturbation to break the integrability.

The purpose of our work is to elucidate possible consequences of the three-body interaction for an energy transport of a Bose gas in one dimension by studying its thermal conductivity in the weak-coupling limit.
To this end, we first formulate our system in Sec.~\ref{sec:preliminary} as well as present its basic properties necessary for later analyses.
Then, the thermal conductivity is evaluated microscopically with the Kubo formula in Sec.~\ref{sec:microscopic} and also based on the quantum Boltzmann equation in Sec.~\ref{sec:kinetic}.
We will find that these two approaches lead to the identical result, where the thermal conductivity is dominated by the three-body interaction rather than the two-body interaction.
Finally, its temperature dependence is numerically determined in Sec.~\ref{sec:thermal} and our findings are summarized in Sec.~\ref{sec:summary}.
Appendix~\ref{app:resummation} presents complementary information regarding the frequency-dependent thermal conductivity and the approximate resummation scheme.

Our work partly follows the analysis described in Ref.~\cite{Fujii:2021} by adapting it for a Bose gas in one dimension with weak two-body and three-body interactions but at arbitrary temperature.
We will set $\hbar=k_B=1$ throughout this paper and an integration over wave number or momentum is denoted by $\int_p\equiv\int_\R dp/(2\pi)$ for the sake of brevity.

\section{Preliminaries}\label{sec:preliminary}
\subsection{Hamiltonian and conservation laws}
Let us consider a Bose gas in quasi-one-dimension realized by confining weakly interacting bosons with a two-dimensional harmonic potential.
Its Hamiltonian density for $\hat\Phi\equiv\hat\Phi(\r)$ is provided by
\begin{align}
\hat\H_\mathrm{3D} = \frac{\grad\hat\Phi^\+\cdot\grad\hat\Phi}{2m}
+ \frac{y^2+z^2}{2ml_\perp^4}\hat\Phi^\+\hat\Phi
+ \frac{g_\mathrm{3D}}{2}\hat\Phi^\+\hat\Phi^\+\hat\Phi\hat\Phi,
\end{align}
where $l_\perp\equiv1/\sqrt{m\omega_\perp}$ is the harmonic oscillator length and $g_\mathrm{3D}\equiv4\pi a_\mathrm{3D}/m$ is the two-body coupling in free space.
As far as low-energy physics relative to the transverse excitation energy is concerned, i.e., $\N,\sqrt{mT}\ll1/l_\perp$, our system is effectively described by the one-dimensional Hamiltonian density of
\begin{align}
\hat\H = \frac{\d_x\hat\phi^\+\d_x\hat\phi}{2m}
+ \frac{g_2}{2}(\hat\phi^\+\hat\phi)^2 + \frac{g_3}{6}(\hat\phi^\+\hat\phi)^3 + \cdots.
\end{align}
Here, the normal ordering is implicitly understood for $\hat\phi\equiv\hat\phi(x)$, $g_2$ and $g_3$ are two-body and three-body couplings, respectively, and the dots include higher $N$-body interactions with $g_{N\geq4}$ as well as interactions involving derivatives.
The effective one-dimensional couplings for $|a_\mathrm{3D}|\ll l_\perp$ are found to be
\begin{align}\label{eq:coupling}
g_2 = 2\frac{a_\mathrm{3D}}{ml_\perp^2}, \qquad
g_3 = -12\ln(4/3)\frac{a_\mathrm{3D}^2}{ml_\perp^2},
\end{align}
and $g_N\sim a_\mathrm{3D}^{N-1}/(ml_\perp^2)$ to the leading orders in $a_\mathrm{3D}$~\cite{Olshanii:1998,Tan:2010,Mazets:2010,Nishida:2018}.%
\footnote{Our expression for $g_3$ agrees with Refs.~\cite{Tan:2010,Mazets:2010,Nishida:2018} but is four times smaller than the earlier one presented in Refs.~\cite{Sinha:2006,Mazets:2008}.
For the sake of clarification, we note that ours is obtained as a result of $g_3=6\times\sum_{n=1}^\infty[g_\mathrm{3D}\int_0^\infty\!dr\,2\pi r\,\psi_0(r)^3\psi_n(r)]^2/(E_0-E_n)$, where $\psi_n(r)=e^{-(r/l_\perp)^2/2}L_n[(r/l_\perp)^2]/(\sqrt\pi\,l_\perp)$ is the isotropic wave function in a two-dimensional harmonic potential and $E_n=(2n+1)\,\omega_\perp$ is its energy eigenvalue.}
Although such multibody interactions arising from virtual transverse excitations are usually neglected in a dilute gas, the thermal conductivity in one dimension will turn out to be dominated by the three-body interaction rather than the two-body interaction.
Therefore, it is necessary and sufficient for our purpose to keep up to the three-body coupling and regard $g_2\sim g_3\sim O(g)$ as small perturbations.

The particle number, momentum, and energy are conserved in our system.
With the help of the equation of motion,
\begin{align}
i\d_t\hat\phi = -\frac{\d_x^2\hat\phi}{2m}
+ g_2\hat\phi^\+\hat\phi^2 + \frac{g_3}{2}\hat\phi^{\+2}\hat\phi^3,
\end{align}
the number density $\hat\N=\hat\phi^\+\hat\phi$, the number flux $\hat\J=[\hat\phi^\+(\d_x\hat\phi)-(\d_x\hat\phi^\+)\hat\phi]/(2im)$, and the Hamiltonian density are found to obey the following continuity equations,
\begin{align}
\d_t\hat\N + \d_x\hat\J &= 0, \\
m\d_t\hat\J + \d_x\hat\Pi &= 0, \\
\d_t\hat\H + \d_x\hat\K &= 0,
\end{align}
where
\begin{align}
\hat\Pi = \frac{\d_x\hat\phi^\+\d_x\hat\phi}{m} - \frac{\d_x^2(\hat\phi^\+\hat\phi)}{4m}
+ \frac{g_2}{2}(\hat\phi^\+\hat\phi)^2 + \frac{g_3}{3}(\hat\phi^\+\hat\phi)^3
\end{align}
is the stress tensor (simply a force in one dimension) and
\begin{align}\label{eq:energy-flux}
\hat\K = \left(\frac{\d_x\hat\phi^\+\d_x^2\hat\phi}{4im^2}
+ g_2\frac{\hat\phi^{\+2}\d_x\hat\phi^2}{4im}
+ g_3\frac{\hat\phi^{\+3}\d_x\hat\phi^3}{12im}\right) + \mathrm{H.c.}
\end{align}
is the energy flux.

\begin{figure}[t]
\includegraphics[width=\columnwidth]{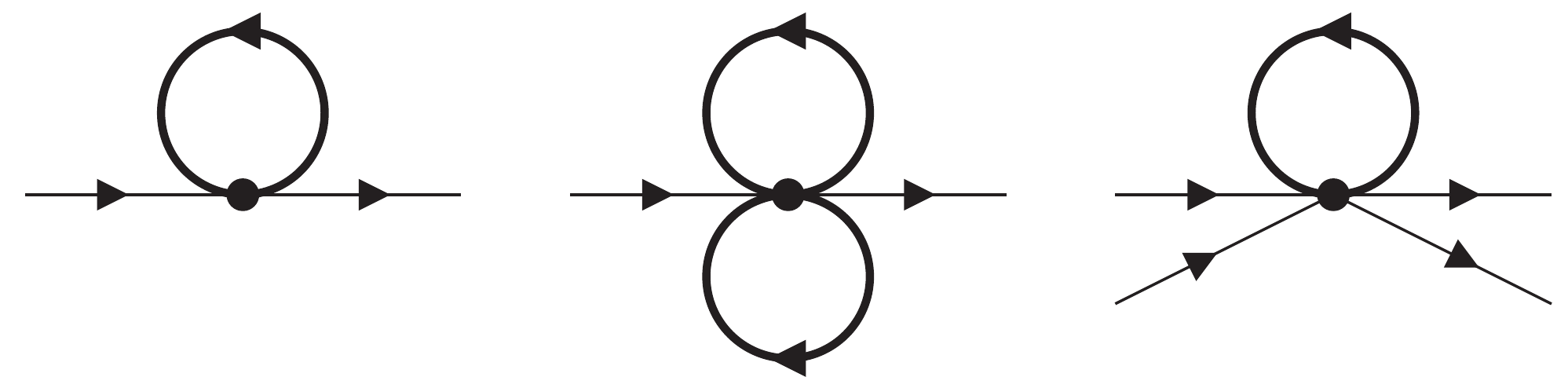}
\caption{\label{fig:mean-field}
Mean-field corrections to the chemical potential (left and middle diagrams) and the two-body coupling (right diagram) in Eq.~(\ref{eq:mean-field}).
The thin and thick lines are the bare and full boson propagators, respectively, whereas the dot is either the two-body or three-body coupling.
The thick loop provides the number density.}
\end{figure}

\subsection{Self-energy and transition rates}
\begin{figure}[t]
\includegraphics[width=\columnwidth]{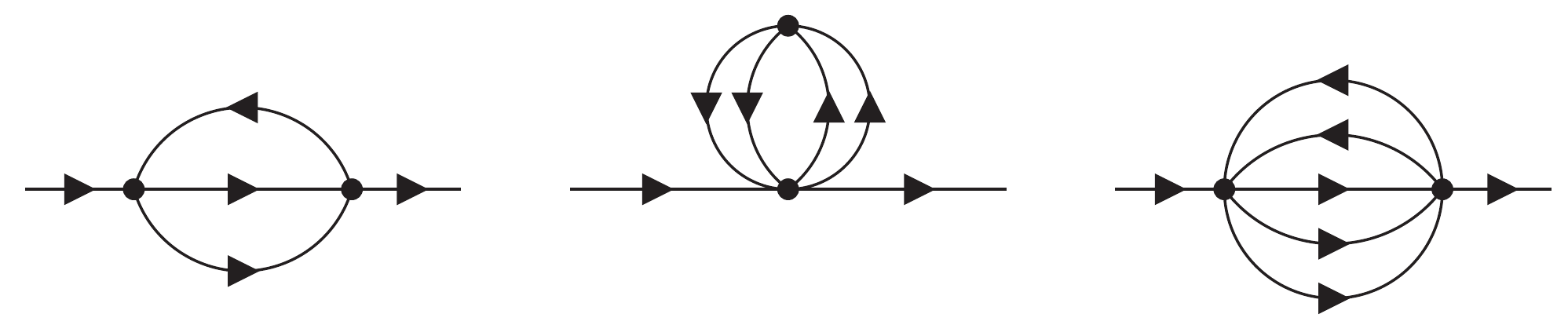}
\caption{\label{fig:self-energy}
Boson self-energies up to $O(g^2)$ corresponding to $\Sigma_{22}$ (left diagram), $\Sigma_{23}$ (middle diagram), and $\Sigma_{33}$ (right diagram) in Eqs.~(\ref{eq:Sigma22})--(\ref{eq:Sigma33}), where the two-body coupling is understood to be $\g_2$.}
\end{figure}

We mostly work with the Matsubara formalism at inverse temperature $\beta=1/T$ and chemical potential $\mu<0$, where the bare boson propagator in the Fourier space is denoted by
\begin{align}
G(iw,p) = \frac1{iw-\ep+\mu}
\end{align}
and the full boson propagator by
\begin{align}
\G(iw,p) = \frac1{iw-\ep+\bar\mu-\Sigma(iw,p)}.
\end{align}
Here, $w=2\pi n/\beta$ with $n\in\Z$ is the bosonic frequency, $\ep=p^2/(2m)$ is the single-particle energy, and $\Sigma(iw,p)$ is the boson self-energy.
Furthermore, it is convenient for diagrammatic calculations to employ
\begin{align}\label{eq:mean-field}
\bar\mu \equiv \mu - 2g_2\N - 3g_3\N^2, \qquad
\g_2 \equiv g_2 + 3g_3\N,
\end{align}
so as to incorporate the mean-field corrections from the number density $\N=\<\hat\phi^\+\hat\phi\>$ as depicted in Fig.~\ref{fig:mean-field}.
Consequently, diagrams involving self-contracted vertices can be eliminated and there remain three self-energy diagrams up to $O(g^2)$, which are depicted in Fig.~\ref{fig:self-energy} and read
\begin{align}\label{eq:Sigma22}
\Sigma_{22}(iw,p) &= \frac{2\g_2^2}{\beta^2}\sum_{w',v}\int_{p',q}
G(iw',p')G(iv,q) \notag\\
&\quad \times G(iw+iw'-iv,p+p'-q),
\end{align}
\begin{align}
\Sigma_{23} &= -\frac{3\g_2g_3}{\beta^3}\sum_{w',w'',v'}\int_{p',p'',q'}
G(iw',p')G(iw'',p'') \notag\\
&\quad \times G(iv',q')G(iw'+iw''-iv',p'+p''-q'),
\end{align}
\begin{align}\label{eq:Sigma33}
& \Sigma_{33}(iw,p) = \frac{3g_3^2}{\beta^4}\sum_{w',w'',v,v'}\int_{p',p'',q,q'} \notag\\
&\quad \times G(iw',p')G(iw'',p'')G(iv,q)G(iv',q') \notag\\
&\quad \times G(iw+iw'+iw''-iv-iv',p+p'+p''-q-q').
\end{align}
Obviously, $\Sigma_{23}$ is independent of $(iw,p)$ and real.

The Matsubara frequency summations in $\Sigma(iw,p)=\Sigma_{22}(iw,p)+\Sigma_{23}+\Sigma_{33}(iw,p)+O(g^3)$ are readily performed by replacing them with the complex contour integrations and the resulting expressions can be arranged into
\begin{align}
& \Sigma_{22}(iw,p) = 2\g_2^2\int_{p',q,q'}
\frac{2\pi\delta(p+p'-q-q')}{iw+\eps_{p'}-\eq-\eps_{q'}+\mu} \notag\\
&\quad \times [f_{p'}(1+f_q)(1+f_{q'}) - (1+f_{p'})f_qf_{q'}],
\end{align}
\begin{align}
& \Sigma_{23} = 3\g_2g_3\int_{p',p'',q',q''}
\frac{2\pi\delta(p'+p''-q'-q'')}{\eps_{p'}+\eps_{p''}-\eps_{q'}-\eps_{q''}} \notag\\
&\quad \times [f_{p'}f_{p''}(1+f_{q'})(1+f_{q''}) - (1+f_{p'})(1+f_{p''})f_{q'}f_{q''}],
\end{align}
\begin{align}
& \Sigma_{33}(iw,p) \notag\\
&= 3g_3^2\int_{p',p'',q,q',q''}\frac{2\pi\delta(p+p'+p''-q-q'-q'')}
{iw+\eps_{p'}+\eps_{p''}-\eq-\eps_{q'}-\eps_{q''}+\mu} \notag\\
&\quad \times [f_{p'}f_{p''}(1+f_q)(1+f_{q'})(1+f_{q''}) \notag\\
&\qquad - (1+f_{p'})(1+f_{p''})f_qf_{q'}f_{q''}],
\end{align}
where $f_p=1/[e^{\beta(\ep-\mu)}-1]$ is the Bose-Einstein distribution function.
From the imaginary part of the on-shell self-energy, $-2\Im[\Sigma(\ep-\mu+i0^+,p)]$, interpreted as the width due to scatterings, the two-body and three-body transition rates from initial momenta $p,p'(,p'')$ to final momenta $q,q'(,q'')$ are identified as
\begin{align}\label{eq:vacuum_2-body}
W_2(p,p'|q,q')
&= 2\g_2^2\,(2\pi)^2\delta(\ep+\eps_{p'}-\eq-\eps_{q'}) \notag\\
&\quad \times \delta(p+p'-q-q')
\end{align}
and
\begin{align}\label{eq:vacuum_3-body}
& W_3(p,p',p''|q,q',q'') \notag\\
&= 3g_3^2\,(2\pi)^2\delta(\ep+\eps_{p'}+\eps_{p''}-\eq-\eps_{q'}-\eps_{q''}) \notag\\
&\quad \times \delta(p+p'+p''-q-q'-q''),
\end{align}
respectively, to the lowest order in perturbation.
Furthermore, it is convenient for later use to introduce the in-medium transition rates by multiplying the statistical weights according to
\begin{align}\label{eq:medium_2-body}
\W_2(p;p'|q,q') = W_2(p,p'|q,q')\,\frac{f_{p'}(1+f_q)(1+f_{q'})}{1+f_p}
\end{align}
and
\begin{align}\label{eq:medium_3-body}
& \W_3(p;p',p''|q,q',q'') = W_3(p,p',p''|q,q',q'') \notag\\
&\quad \times \frac{f_{p'}f_{p''}(1+f_q)(1+f_{q'})(1+f_{q''})}{1+f_p},
\end{align}
so that the imaginary part of the on-shell self-energy is simply expressed as
\begin{align}\label{eq:self-energy}
& -2\Im[\Sigma(\ep-\mu+i0^+,p)] = \int_{p',q,q'}\W_2(p;p'|q,q') \notag\\
&\quad + \int_{p',p'',q,q',q''}\W_3(p;p',p''|q,q',q'').
\end{align}
Here, $f_pf_{p'}(1+f_q)(1+f_{q'})=(1+f_p)(1+f_{p'})f_qf_{q'}$ under $\ep+\eps_{p'}=\eq+\eps_{q'}$ and its three-body generalization are employed.

\section{Microscopic theory}\label{sec:microscopic}
\subsection{Kubo formula and pinch singularity}
The thermal conductivity is the transport coefficient relating the heat flux to the temperature gradient.
According to the linear-response theory, it is microscopically provided by the Kubo formula~\cite{Kubo:1957a,Kubo:1957b},
\begin{align}\label{eq:kubo}
T\kappa = \lim_{\omega\to0}\frac{\Im[\chi_Q(\omega+i0^+)]}{\omega},
\end{align}
where $\chi_Q(\omega+i0^+)$ is the retarded correlation function at zero wave number and is most conveniently obtained from the corresponding imaginary-time-ordered correlation function,
\begin{align}\label{eq:correlation}
\chi_Q(iw) = \frac1L\int_0^\beta\!d\tau\,e^{iw\tau}\<\T\,\hat{Q}(\tau)\hat{Q}(0)\>,
\end{align}
with an analytic continuation of $iw\to\omega+i0^+$~\cite{Altland-Simons}.
Here, $L=\int dx$ is the system size and $\hat{Q}=\int dx\,[\hat\K-(\E+\P)\hat\J/\N]$ is the integrated heat flux operator with $\N$, $\E$, and $\P$ being the number density, the energy density, and the pressure, respectively~\cite{Mori:1962,Kadanoff:1963,Luttinger:1964}.
As far as the lowest order in perturbation is concerned, the two-body and three-body operators in Eq.~(\ref{eq:energy-flux}) are irrelevant, leading to
\begin{align}\label{eq:operator}
\hat{Q} = \int_p\Q_p\,\hat\phi_p^\+\hat\phi_p + O(g),
\end{align}
where $\hat\phi_p=\int dx\,e^{-ipx}\hat\phi(x)$ is a Fourier component of the field operator and
\begin{align}\label{eq:heat-flux}
\Q_p = \left(\ep-\frac{\E+\P}{\N}\right)\frac{p}{m}
\end{align}
is the single-particle heat flux serving as the bare vertex function for the thermal conductivity.

\begin{figure}[t]
\includegraphics[width=0.4\columnwidth]{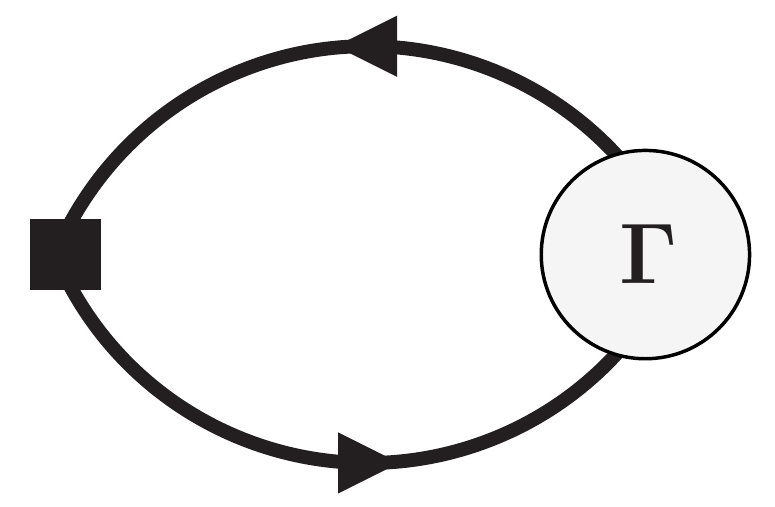}
\caption{\label{fig:correlation}
Diagrammatic representation of the imaginary-time-ordered correlation function in Eq.~(\ref{eq:time-ordered}).
The square and the circle ($\Gamma$) are the bare and full vertex functions, respectively.}
\end{figure}

For the one-body operator in the form of Eq.~(\ref{eq:operator}), the imaginary-time-ordered correlation function can formally be expressed as
\begin{align}\label{eq:time-ordered}
\chi_Q(iw) &= \frac1\beta\sum_v\int_p\Q_p\,\G(iv+iw,p)\G(iv,p) \notag\\
&\quad \times \Gamma(iv+iw,iv;p),
\end{align}
whose diagrammatic representation is depicted in Fig.~\ref{fig:correlation} with $\Gamma(iv+iw,iv;p)$ being the full vertex function to be specified below.
The Matsubara frequency summation is replaced with the complex contour integration over $iv\to\nu$ and its contour is deformed into four lines along $\Im(\nu)=\pm0^+,-w\pm0^+$ so as to avoid possible singularities of the integrand~\cite{Eliashberg:1962,Fujii:2021}.
Then, the analytic continuation of $iw\to\omega+i0^+$ leads to
\begin{align}\label{eq:retarded}
& \chi_Q(\omega+i0^+) = \int_{\R\setminus\{0\}}\!\frac{d\nu}{2\pi i}\,
\frac1{e^{\beta\nu}-1}\int_p\Q_p \notag\\
& \times \bigl[\G_+(\nu+\omega,p)\G_+(\nu,p)\Gamma(\nu+\omega+i0^+,\nu+i0^+;p) \notag\\
&\quad - \G_+(\nu+\omega,p)\G_-(\nu,p)\Gamma(\nu+\omega+i0^+,\nu-i0^+;p) \notag\\
&\quad + \G_+(\nu,p)\G_-(\nu-\omega,p)\Gamma(\nu+i0^+,\nu-\omega-i0^+;p) \notag\\
&\quad - \G_-(\nu,p)\G_-(\nu-\omega,p)\Gamma(\nu-i0^+,\nu-\omega-i0^+;p)\bigr],
\end{align}
where $\G_\pm(\nu,p)\equiv\G(\nu\pm i0^+,p)$ are the retarded (upper sign) and advanced (lower sign) boson propagators.

Although the retarded correlation function in Eq.~(\ref{eq:retarded}) is apparently $O(g^0)$, such a naive order counting breaks down in the zero-frequency limit, $\omega\to0$, due to the pinch singularity~\cite{Eliashberg:1962,Jeon:1995,Jeon:1996,Hidaka:2011}.
Actually, the product of retarded and advanced boson propagators with the same frequency and wave number is evaluated as
\begin{align}\label{eq:pinch}
\G_+(\nu,p)\G_-(\nu,p)
= -\frac{\pi\,\delta(\nu-\ep+\mu)}{\Im[\Sigma(\nu+i0^+,p)]} + O(g^{-1}),
\end{align}
giving rise to the inverse square order in perturbation because of $\Im[\G(\nu+i0^+,p)]=-\pi\,\delta(\nu-\ep+\mu)+O(g)$ and $\Im[\Sigma(\nu+i0^+,p)]\sim O(g^2)$ in the weak-coupling limit.
Therefore, the resulting thermal conductivity from Eq.~(\ref{eq:kubo}) involves an $O(g^{-2})$ contribution provided by
\begin{align}\label{eq:thermal}
\kappa = \beta^2\int_p f_p(1+f_p)\Q_p
\frac{\Gamma_{+-}(p)}{-2\Im[\Sigma_+(p)]} + O(g^{-1}).
\end{align}
Here, shorthand notations for on-shell $\Sigma_+(p)\equiv\Sigma(\ep-\mu+i0^+,p)$ and $\Gamma_{+-}(p)\equiv\Gamma(\ep-\mu+i0^+,\ep-\mu-i0^+;p)$ are introduced and the latter is to be $O(g^0)$ and real because of $[\Gamma(w,v;p)]^*=\Gamma(v^*,w^*;p)$.

\begin{figure}[t]
\includegraphics[width=0.7\columnwidth]{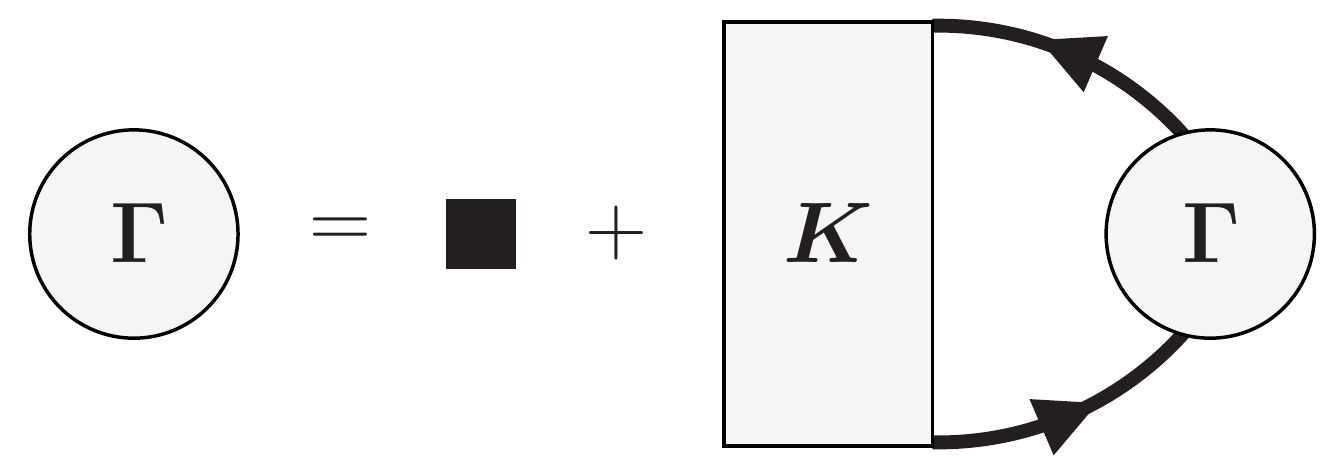}
\caption{\label{fig:vertex}
Diagrammatic representation of the self-consistent equation for the full vertex function in Eq.~(\ref{eq:self-consistent}), which incorporates an iteration of four-point functions represented by the rectangle ($K$).}
\end{figure}

\subsection{Vertex self-consistent equation}
\begin{figure}[t]
\includegraphics[width=0.7\columnwidth]{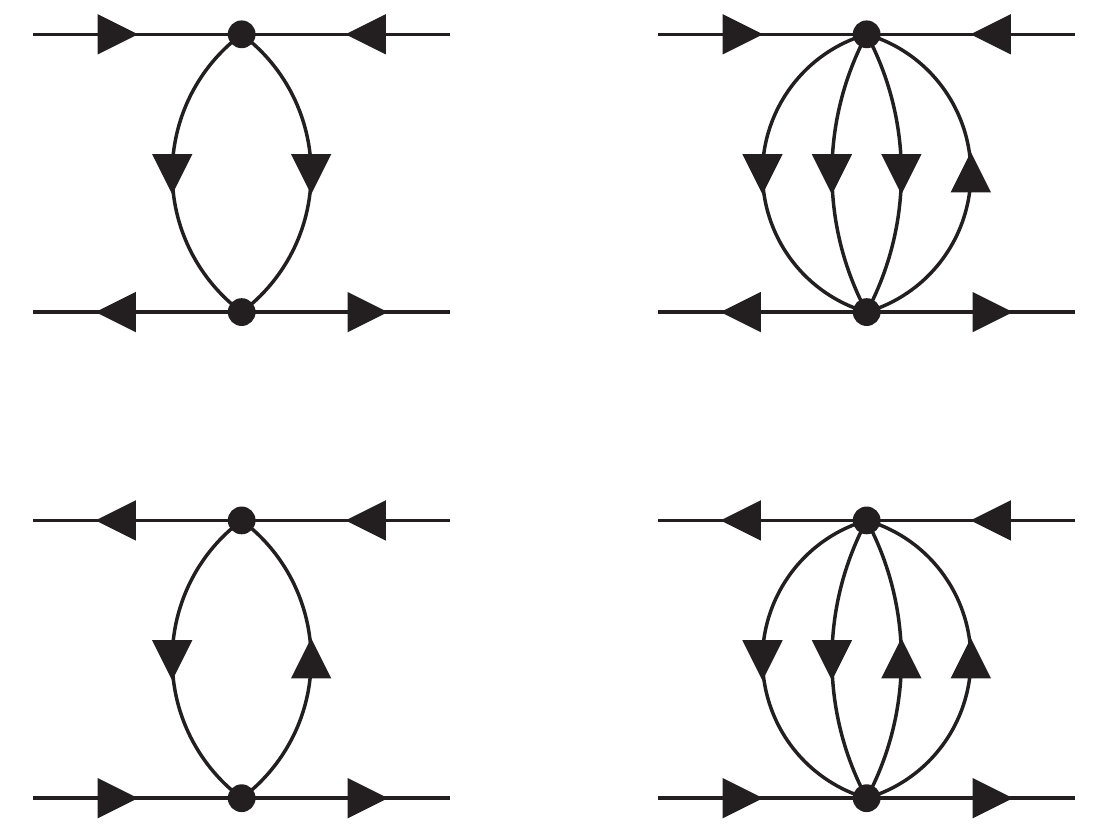}
\caption{\label{fig:kernel}
Four-point functions up to $O(g^2)$ corresponding to $K^\MT_{22}$ (upper left diagram), $K^\MT_{33}$ (upper right diagram), $K^\AL_{22}$ (lower left diagram), and $K^\AL_{33}$ (lower right diagram) in Eqs.~(\ref{eq:MT22_def})--(\ref{eq:AL33_def}).}
\end{figure}

The full vertex function is provided by summing up an infinite sequence of diagrams which is organized into an iteration of four-point functions connected by a pair of counterpropagating boson propagators~\cite{Fujii:2021}.
Such a summation can formally be achieved by imposing the following self-consistent equation on the vertex function,
\begin{align}\label{eq:self-consistent}
& \Gamma(iv+iw,iv;p) \notag\\
&= \Q_p + \frac1\beta\sum_{v'}\int_{p'}K(iv+iw,iv;p|iv'+iw,iv';p') \notag\\
&\quad \times \G(iv'+iw,p')\G(iv',p')\Gamma(iv'+iw,iv';p'),
\end{align}
whose diagrammatic representation is depicted in Fig.~\ref{fig:vertex}.
Here, $K(iv+iw,iv;p|iv'+iw,iv';p')$ is the four-point function, which to the lowest order in perturbation is $O(g^2)$ and consists of four diagrams depicted in Fig.~\ref{fig:kernel}.%
\footnote{Although there is an additional $O(g)$ diagram contributing $K(\,*\,)=2\g_2$, it vanishes when substituted into Eq.~(\ref{eq:self-consistent}) because the integrand turns odd in $p'$.}
They read
\begin{align}\label{eq:MT22_def}
& K^\MT_{22}(iv+iw,iv;p|iv'+iw,iv';p') \notag\\
&= \frac{2\g_2^2}{\beta}\sum_u\int_qG(iu,q)G(iv+iv'+iw-iu,p+p'-q),
\end{align}
\begin{align}
& K^\MT_{33}(iv+iw,iv;p|iv'+iw,iv';p') \notag\\
&= \frac{6g_3^2}{\beta^3}\sum_{v'',u,u'}\int_{p'',q,q'}
G(iv'',p'')G(iu,q)G(iu',q') \notag\\
&\quad \times G(iv+iv'+iv''+iw-iu-iu',p+p'+p''-q-q')
\end{align}
for the type of Maki-Thompson and
\begin{align}
& K^\AL_{22}(iv+iw,iv;p|iv'+iw,iv';p') \notag\\
&= \frac{4\g_2^2}{\beta}\sum_u\int_qG(iu,q)G(iv-iv'+iu,p-p'+q),
\end{align}
\begin{align}\label{eq:AL33_def}
& K^\AL_{33}(iv+iw,iv;p|iv'+iw,iv';p') \notag\\
&= \frac{9g_3^2}{\beta^3}\sum_{v'',u,u'}\int_{p'',q,q'}
G(iv'',p'')G(iu,q)G(iu',q')\notag\\
&\quad \times G(iv-iv'+iv''+iu-iu',p-p'+p''+q-q')
\end{align}
for the type of Aslamazov-Larkin.
The Matsubara frequency summations in $K(\,*\,)=K^\MT_{22}(\,*\,)+K^\MT_{33}(\,*\,)+K^\AL_{22}(\,*\,)+K^\AL_{33}(\,*\,)+O(g^3)$ are readily performed by replacing them with the complex contour integrations and the resulting expressions can be arranged into
\begin{align}\label{eq:MT22_pole}
& K^\MT_{22}(iv+iw,iv;p|iv'+iw,iv';p') \notag\\
&= -2\g_2^2\int_{q,q'}
\frac{2\pi\delta(p+p'-q-q')}{iv+iv'+iw-\eq-\eps_{q'}+2\mu} \notag\\
&\quad \times [(1+f_q)(1+f_{q'}) - f_qf_{q'}],
\end{align}
\begin{align}
& K^\MT_{33}(iv+iw,iv;p|iv'+iw,iv';p') \notag\\
&= -6g_3^2\int_{p'',q,q',q''}\frac{2\pi\delta(p+p'+p''-q-q'-q'')}
{iv+iv'+iw+\eps_{p''}-\eq-\eps_{q'}-\eps_{q''}+2\mu} \notag\\
&\quad \times [f_{p''}(1+f_q)(1+f_{q'})(1+f_{q''}) - (1+f_{p''})f_qf_{q'}f_{q''}],
\end{align}
as well as
\begin{align}
& K^\AL_{22}(iv+iw,iv;p|iv'+iw,iv';p') \notag\\
&= -4\g_2^2\int_{q,q'}
\frac{2\pi\delta(p-p'+q-q')}{iv-iv'+\eq-\eps_{q'}} \notag\\
&\quad \times [f_q(1+f_{q'}) - (1+f_q)f_{q'}],
\end{align}
\begin{align}\label{eq:AL33_pole}
& K^\AL_{33}(iv+iw,iv;p|iv'+iw,iv';p') \notag\\
&= -9g_3^2\int_{p'',q,q',q''}\frac{2\pi\delta(p-p'+p''+q-q'-q'')}
{iv-iv'+\eps_{p''}+\eq-\eps_{q'}-\eps_{q''}} \notag\\
&\quad \times [f_{p''}f_q(1+f_{q'})(1+f_{q''}) - (1+f_{p''})(1+f_q)f_{q'}f_{q''}].
\end{align}

By substituting these four-point functions into Eq.~(\ref{eq:self-consistent}), we obtain
\begin{align}\label{eq:vertex}
& \Gamma(iv+iw,iv;p) \notag\\
&= \Q_p + \Gamma^\MT_{22}(iv+iw,iv;p) + \Gamma^\MT_{33}(iv+iw,iv;p) \notag\\
&\quad + \Gamma^\AL_{22}(iv+iw,iv;p) + \Gamma^\AL_{33}(iv+iw,iv;p) + O(g^3)
\end{align}
with the individual contribution denoted by
\begin{align}
& \Gamma^{\MT,\AL}_{22,33}(iv+iw,iv;p) \notag\\
&\equiv \frac1\beta\sum_{v'}\int_{p'}K^{\MT,\AL}_{22,33}(iv+iw,iv;p|iv'+iw,iv';p') \notag\\
&\quad \times \G(iv'+iw,p')\G(iv',p')\Gamma(iv'+iw,iv';p').
\end{align}
Each Matsubara frequency summation is replaced with the complex contour integration over $iv'\to\nu'$, whose integrand has possible singularities only along $\Im(\nu')=0,-w$ in addition to the simple pole explicit in Eqs.~(\ref{eq:MT22_pole})--(\ref{eq:AL33_pole})~\cite{Eliashberg:1962,Fujii:2021}.
Therefore, its contour is deformed into four lines along $\Im(\nu)=\pm0^+,-w\pm0^+$ and one circle around the pole, whereas only three of them along $\Im(\nu')=-0^+,-w+0^+$ and around the pole turn out to suffer from the pinch singularity so as to contribute $O(g^0)$ in the zero-frequency limit.
Then, as far as the lowest order in perturbation is concerned, the analytic continuation of $iv\to\ep-\mu-i0^+$ followed by $iw\to i0^+$ in Eq.~(\ref{eq:vertex}) leads to
\begin{align}
& \Gamma^\MT_{22}(\ep-\mu+i0^+,\ep-\mu-i0^+;p) \notag\\
&= -\int_{p',q,q'}\W_2(p;p'|q,q')\frac{\Gamma_{+-}(p')}{-2\Im[\Sigma_+(p')]} + O(g),
\end{align}
\begin{align}
& \Gamma^\MT_{33}(\ep-\mu+i0^+,\ep-\mu-i0^+;p) \notag\\
&= -2\int_{p',p'',q,q',q''}\W_3(p;p',p''|q,q',q'')
\frac{\Gamma_{+-}(p')}{-2\Im[\Sigma_+(p')]} \notag\\
&\quad + O(g)
\end{align}
for the type of Maki-Thompson and
\begin{align}
& \Gamma^\AL_{22}(\ep-\mu+i0^+,\ep-\mu-i0^+;p) \notag\\
&= 2\int_{p',q,q'}\W_2(p;p'|q,q')\frac{\Gamma_{+-}(q)}{-2\Im[\Sigma_+(q)]} + O(g),
\end{align}
\begin{align}
& \Gamma^\AL_{33}(\ep-\mu+i0^+,\ep-\mu-i0^+;p) \notag\\
&= 3\int_{p',p'',q,q',q''}\W_3(p;p',p''|q,q',q'')
\frac{\Gamma_{+-}(q)}{-2\Im[\Sigma_+(q)]} \notag\\
&\quad + O(g)
\end{align}
for the type of Aslamazov-Larkin.
Here, Eq.~(\ref{eq:pinch}) is applied for the pinch singularity, as well as Eqs.~(\ref{eq:medium_2-body}) and (\ref{eq:medium_3-body}) for the in-medium transition rates, and the integration variables are exchanged as $p'\leftrightarrow q$ in $\Gamma^\AL_{22,33}$.

Consequently, we find that the self-consistent equation for the vertex function is analytically continued into the following on-shell equation,
\begin{align}\label{eq:on-shell}
\Q_p &= \int_{p',q,q'}\W_2(p;p'|q,q')\,
(\varphi_p + \varphi_{p'} - \varphi_q - \varphi_{q'}) \notag\\
&\quad + \int_{p',p'',q,q',q''}\W_3(p;p',p''|q,q',q'') \notag\\
&\qquad \times (\varphi_p + \varphi_{p'} + \varphi_{p''}
- \varphi_q - \varphi_{q'} - \varphi_{q''}) + O(g),
\end{align}
where $\varphi_p\equiv\Gamma_{+-}(p)/[-2\Im[\Sigma_+(p)]]$ is introduced with the imaginary part of the on-shell self-energy provided by Eq.~(\ref{eq:self-energy})~\cite{Fujii:2021}.
Once the solution of the resulting integral equation closed for $\varphi_p\sim O(g^{-2})$ is obtained, the thermal conductivity in Eq.~(\ref{eq:thermal}) is provided by
\begin{align}\label{eq:conductivity}
\kappa = \beta^2\int_pf_p(1+f_p)\Q_p\varphi_p + O(g^{-1})
\end{align}
to the lowest order in perturbation.
Importantly, the energy and momentum conservations in $\W_2(p;p'|q,q')$ equate $\{p,p'\}$ and $\{q,q'\}$ so that the set of two momenta is unchangeable by the two-body scattering.
Therefore, the first term on the right-hand side of Eq.~(\ref{eq:on-shell}) vanishes identically and the thermal conductivity in one dimension is indeed dominated by the three-body interaction rather than the two-body interaction.

\section{Kinetic theory}\label{sec:kinetic}
\subsection{Quantum Boltzmann equation}
An alternative approach to compute the thermal conductivity is based on the quantum Boltzmann equation~\cite{Lifshitz-Pitaevskii}:
\begin{align}\label{eq:boltzmann}
\frac{\d F_p}{\d t} + \frac{\d\ep}{\d p}\frac{\d F_p}{\d x}
= \left(\frac{\d F_p}{\d t}\right)_\cl.
\end{align}
Here, $F_p=F_p(t,x)$ is a local distribution function of bosons and their two-body and three-body interactions lead to the collision term of
\begin{align}
& \left(\frac{\d F_p}{\d t}\right)_\cl
= \int_{p',q,q'}W_2(p,p'|q,q') \notag\\
&\quad \times [(1+F_p)(1+F_{p'})F_qF_{q'}-F_pF_{p'}(1+F_q)(1+F_{q'})] \notag\\
& + \int_{p',p'',q,q',q''}W_3(p,p',p''|q,q',q'') \notag\\
&\quad \times [(1+F_p)(1+F_{p'})(1+F_{p''})F_qF_{q'}F_{q''} \notag\\
&\qquad - F_pF_{p'}F_{p''}(1+F_q)(1+F_{q'})(1+F_{q''})],
\end{align}
with the transition rates provided by Eqs.~(\ref{eq:vacuum_2-body}) and (\ref{eq:vacuum_3-body}) for our system.
However, because of $\{p,p'\}=\{q,q'\}$ imposed by the energy and momentum conservations in $W_2(p,p'|q,q')$, the first term on the right-hand side vanishes identically so that the collision term in one dimension is again dominated by the three-body interaction rather than the two-body interaction.

According to the Chapman-Enskog expansion, the distribution function is decomposed into $F_p=\f_p+\delta f_p$, where
\begin{align}
\f_p \equiv f_{p-mv} = \frac1{\exp[\beta(\eps_{p-mv}-\mu)] - 1}
\end{align}
is a local equilibrium distribution function of bosons and the deviation $\delta f_p$ is assumed to satisfy
\begin{align}\label{eq:condition}
\int_p\delta f_p = \int_pp\,\delta f_p = \int_p\ep\delta f_p = 0.
\end{align}
Therefore, the local chemical potential $\mu=\mu(t,x)$, the local velocity $v=v(t,x)$, and the local inverse temperature $\beta=\beta(t,x)$ are determined so that $\f_p$ solely constitutes local number, local momentum, and local energy densities.
In particular, the number and energy densities in the rest frame are provided by
\begin{align}\label{eq:number}
\N = \int_pf_p = \sqrt{\frac{m}{2\pi\beta}}\Li_{1/2}(e^{\beta\mu})
\end{align}
and
\begin{align}\label{eq:energy}
\E = \int_p\ep f_p = \frac1{2\beta}\sqrt{\frac{m}{2\pi\beta}}\Li_{3/2}(e^{\beta\mu}),
\end{align}
respectively, whereas the pressure obeys the ideal gas law of $\P=2\E$.

From the continuity equations derived with the Boltzmann equation, the stress tensor and the energy flux are respectively identified as
\begin{align}
\Pi = \int_p\frac{p^2}{m}F_p, \qquad
\K = \int_p\ep\frac{p}{m}F_p.
\end{align}
Then, the substitution of $F_p=\f_p+\delta f_p$ brings the former into $\Pi=\P+m\N v^2$ with no dissipative correction in one dimension, whereas the latter is decomposed into $\K=[\eps_{mv}+(\E+\P)/\N]\N v+k$ with the dissipative correction of
\begin{align}\label{eq:dissipation}
k = \int_p\Q_{p-mv}\,\delta f_p.
\end{align}

\subsection{Linearization}
When the system is slightly out of homogeneous thermodynamic equilibrium, $\delta f_p\ll\f_p$ can be determined by linearizing the quantum Boltzmann equation~\cite{Lifshitz-Pitaevskii}.
The substitution of $F_p=\f_p+\delta f_p$ into the left-hand side (LHS) of Eq.~(\ref{eq:boltzmann}) with the help of the thermodynamic relations and the continuity equations leads to
\begin{align}\label{eq:LHS}
(\LHS) = -\beta\f_p(1+\f_p)\Q_{p-mv}\,\d_x\ln\beta + O(\delta f).
\end{align}
Here, we note that there can be a term proportional to $\d_xv$ but its coefficient is $(p-mv)^2/m-(\d\P/\d\N)_\E-\eps_{p-mv}(\d\P/\d\E)_\N=0$ due to the ideal gas law, indicating the vanishing bulk viscosity within the Boltzmann equation~\cite{Dusling:2013,Chafin:2013,Fujii:2021}.
On the other hand, because the equilibrium distribution function cancels the collision term, the right-hand side (RHS) of Eq.~(\ref{eq:boltzmann}) with $\delta f_p\equiv\beta\f_p(1+\f_p)\phi_p$ introduced is provided by
\begin{align}\label{eq:RHS}
(\RHS) &= -\beta\int_{p',p'',q,q',q''}W_3(p,p',p''|q,q',q'') \notag\\
&\quad \times \f_p\f_{p'}\f_{p''}(1+\f_q)(1+\f_{q'})(1+\f_{q''}) \notag\\
&\quad \times (\phi_p + \phi_{p'} + \phi_{p''}
- \phi_q - \phi_{q'} - \phi_{q''}) + O(\delta f^2).
\end{align}

In order for both sides of the quantum Boltzmann equation in Eqs.~(\ref{eq:LHS}) and (\ref{eq:RHS}) to match for arbitrary $\mu$, $v$, and $\beta$, $\phi_p$ must be in the form of $\phi_p=\varphi_{p-mv}\,\d_x\ln\beta$.
Here, $\varphi_p$ quantifies the deviation from the equilibrium distribution function induced by the temperature gradient and solves
\begin{align}\label{eq:linearized}
\Q_p &= \int_{p',p'',q,q',q''}\W_3(p;p',p''|q,q',q'') \notag\\
&\quad \times (\varphi_p + \varphi_{p'} + \varphi_{p''}
- \varphi_q - \varphi_{q'} - \varphi_{q''}).
\end{align}
Furthermore, the dissipative correction to the energy flux in Eq.~(\ref{eq:dissipation}) turns into
\begin{align}
k = \d_x\beta\int_pf_p(1+f_p)\Q_p\varphi_p,
\end{align}
from which the thermal conductivity defined via $k=-\kappa\,\d_xT$ (Fourier's law) is found to be
\begin{align}\label{eq:formula}
\kappa = \beta^2\int_pf_p(1+f_p)\Q_p\varphi_p.
\end{align}
Therefore, by comparing Eqs.~(\ref{eq:linearized}) and (\ref{eq:formula}) to Eqs.~(\ref{eq:on-shell}) and (\ref{eq:conductivity}), respectively, it is now established that the thermal conductivity computed based on the quantum Boltzmann equation is identical to that computed microscopically with the Kubo formula in the weak-coupling limit.

\section{Thermal conductivity}\label{sec:thermal}
Finally, we solve the linearized Boltzmann equation to compute the thermal conductivity in the weak-coupling limit.
Because $\varphi_p\sim O(g_3^{-2})$ must be an odd function of $p$, we expand it in terms of the Laguerre polynomials as
\begin{align}\label{eq:expansion}
\varphi_p = \frac{p}{m^3g_3^2}\sum_{n'=0}^Nc_{n'}L_{n'}(\beta\ep).
\end{align}
This expansion is truncated up to the $N$th order, where the lowest coefficient $c_0$ is actually indefinite because it disappears from Eq.~(\ref{eq:linearized}) due to the momentum conservation.%
\footnote{From the kinetic theory perspective, $c_0$ is to be fixed so as to satisfy the Chapman-Enskog condition in Eq.~(\ref{eq:condition}).}
However, such arbitrariness does not bring any ambiguity into the thermal conductivity because $c_0$ also disappears from Eq.~(\ref{eq:formula}) due to
\begin{align}\label{eq:integral}
\int_p f_p(1+f_p)\Q_pp = \frac{m}{\beta}\int_p f_p\d_p\Q_p = 0
\end{align}
under Eqs.~(\ref{eq:heat-flux}), (\ref{eq:number}), (\ref{eq:energy}), and $\P=2\E$.
Therefore, $c_0$ can safely be neglected and the other $N$ coefficients are determined by deriving a set of $N$ linear equations from Eq.~(\ref{eq:linearized}), such as
\begin{align}\label{eq:linear}
& \int_pf_p(1+f_p)\Q_ppL_n(\beta\ep) \notag\\
&= \int_{p,p',p'',q,q',q''}W_3(p,p',p''|q,q',q'') \notag\\
&\quad \times f_pf_{p'}f_{p''}(1+f_q)(1+f_{q'})(1+f_{q''}) \notag\\
&\quad \times pL_n(\beta\ep)\,(\varphi_p + \varphi_{p'} + \varphi_{p''}
- \varphi_q - \varphi_{q'} - \varphi_{q''})
\end{align}
for $n=1,2,\dots,N$.%
\footnote{Here, $n=0$ is eliminated because Eq.~(\ref{eq:linear}) is reduced simply to $0=0$.}
It is then convenient to change the integration variables on the right-hand side to $p=\lambda/3+\rho\cos\theta$, $p'=\lambda/3+\rho\cos(\theta+2\pi/3)$, and $p''=\lambda/3+\rho\cos(\theta+4\pi/3)$ with $dp\,dp'dp''=(\sqrt3\rho/2)\,d\lambda\,d\rho\,d\theta$ ($\lambda\in\R$, $\rho\in\R_+$, $\theta\in[0,2\pi]$) and similarly for $q$, $q'$, and $q''$.
Consequently, the energy and momentum conservations in $W_3(p,p',p''|q,q',q'')$ are readily imposed by employing $\ep+\eps_{p'}+\eps_{p''}=\lambda^2/(6m)+3\rho^2/(4m)$ and $p+p'+p''=\lambda$.

\begin{figure}[t]
\includegraphics[width=0.9\columnwidth]{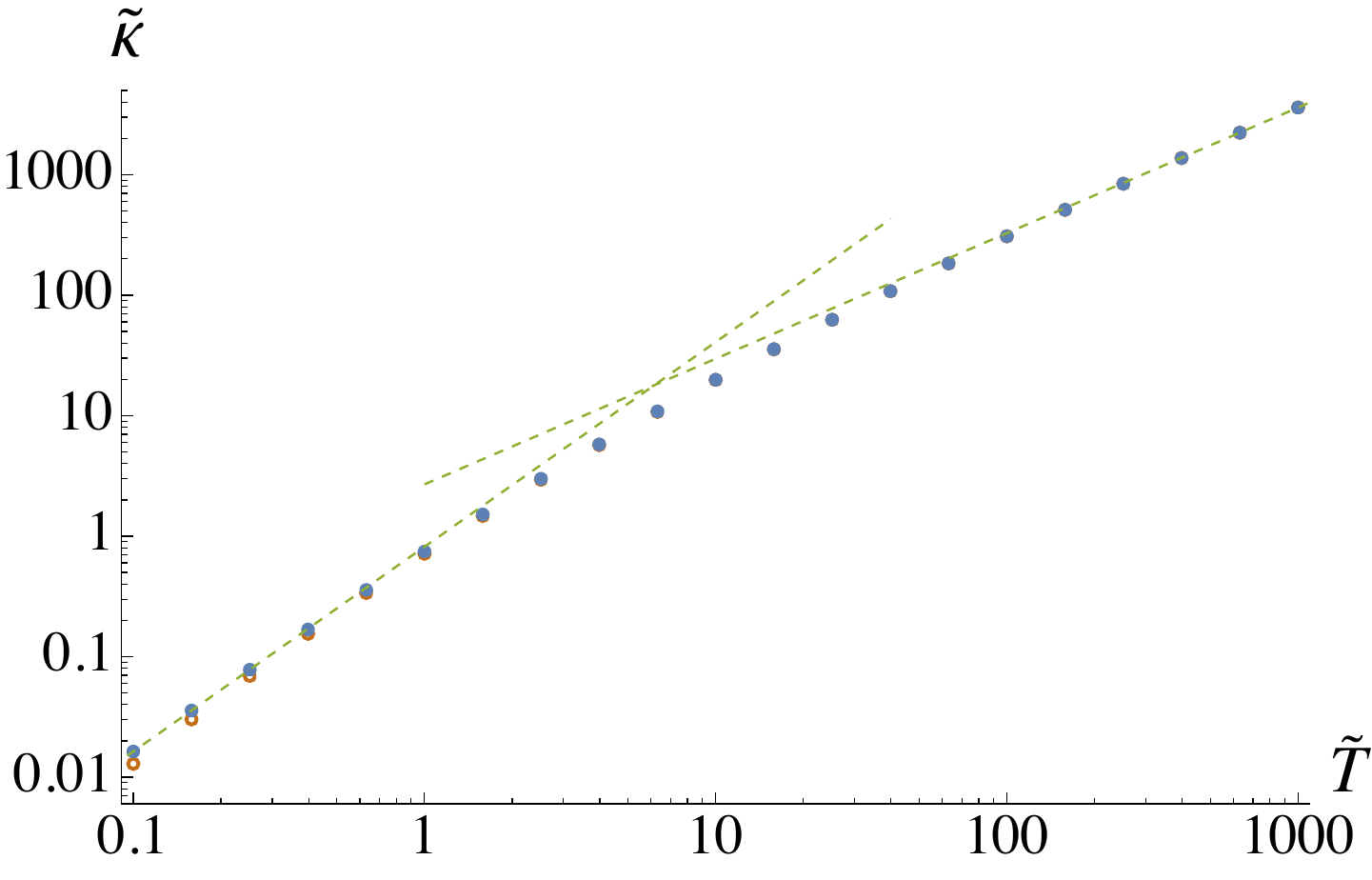}
\caption{\label{fig:kappa-T}
Thermal conductivity normalized by $\tilde\kappa\equiv m^3g_3^2\,\kappa/\N$ in the weak-coupling limit as a function of $\tilde{T}\equiv mT/\N^2$.
The filled dots (blue) indicate the numerically exact result, covering over most of the open dots (red) resulting from the relaxation-time approximation.
The lower left and upper right dashed lines (green) are fits at low and high temperatures, respectively, providing $\tilde\kappa\approx1\times\tilde{T}^{1.7}$ ($\tilde{T}<1$) and $\tilde\kappa\approx3\times\tilde{T}^{1.0}$ ($\tilde{T}>100$).}
\end{figure}

Figure~\ref{fig:kappa-T} shows the thermal conductivity $\kappa$ obtained numerically as a function of the temperature $T$ with the fixed number density $\N$.
Here, $N$ is increased up to $N=16$, which is confirmed to be more than sufficient for convergence of the presented result, so as to make it exact to the lowest order in perturbation.
In particular, we find that $\kappa$ is a monotonically increasing function of $T$ and fits well into the power-law scalings of $\kappa\sim T^{1.7}$ at low temperature and $\kappa\sim T^{1.0}$ at high temperature.

We also note that the linearized Boltzmann equation is often solved with the so-called relaxation-time approximation~\cite{Massignan:2005,Bruun:2005,Braby:2010,Bruun:2012,Schafer:2012}, which identifies the collision term with $(\d F_p/\d t)_\cl=-\delta f_p/\tau$.
This form equated to Eq.~(\ref{eq:LHS}) leads to $\varphi_p=\Q_p\tau$, corresponding to the simplest case of $N=1$ in Eq.~(\ref{eq:expansion}) because the term linear in $p$ is irrelevant according to Eq.~(\ref{eq:integral}).
Then, Eq.~(\ref{eq:linear}) determines the relaxation time as
\begin{align}\label{eq:relaxation}
\frac1\tau &= \frac{\beta^2}{D}\int_{p,p',p'',q,q',q''}W_3(p,p',p''|q,q',q'') \notag\\
&\quad \times f_pf_{p'}f_{p''}(1+f_q)(1+f_{q'})(1+f_{q''}) \notag\\
&\quad \times \Q_p(\Q_p+\Q_{p'}+\Q_{p''}-\Q_q-\Q_{q'}-\Q_{q''})
\end{align}
with $D=\beta^2\int_pf_p(1+f_p)\Q_p^2$ being the Drude weight in the noninteracting limit.%
\footnote{See Eq.~(\ref{eq:kappa_one-loop}) in Appendix~\ref{app:resummation}, where the relaxation-time approximation is described from a different perspective.}
The thermal conductivity resulting from Eq.~(\ref{eq:formula}) reads $\kappa=D\tau$, which is also indicated in Fig.~\ref{fig:kappa-T} for the sake of comparison.
It consistently lies below the numerically exact result and their relative difference is found to increase monotonically by lowering the temperature and exceed 20\% at $mT/\N^2=0.1$.

\section{Summary}\label{sec:summary}
In summary, we studied the thermal conductivity of a weakly interacting Bose gas in one dimension with both two-body and three-body interactions.
Our work is motivated by the fact that an effective three-body interaction inevitably arises as the leading perturbation to break the integrability.
To elucidate its consequences for an energy transport, the Kubo formula was evaluated exactly to the lowest order in perturbation by summing up all contributions that are naively higher orders in perturbation but become comparable in the zero-frequency limit due to the pinch singularity.
Consequently, a self-consistent equation for a vertex function was derived [Eq.~(\ref{eq:on-shell})], which proved to be identical to the linearized version of the quantum Boltzmann equation [Eq.~(\ref{eq:linearized})].
The resulting thermal conductivity was found to be finite, indicating that the energy transport turns diffusive due to the three-body interaction rather than the two-body interaction.

In particular, when our system in quasi-one-dimension is realized by confining weakly interacting bosons with a two-dimensional harmonic potential, the thermal conductivity for $0<a_\mathrm{3D}\ll l_\perp\ll1/\N,1/\sqrt{mT}$ is predicted to be
\begin{align}
\kappa = \frac{(l_\perp/a_\mathrm{3D})^4\N}{[12\ln(4/3)]^2m}\,\tilde\kappa(\tilde T).
\end{align}
Here, Eq.~(\ref{eq:coupling}) is employed and $\tilde\kappa(\tilde T)$ is the dimensionless function obtained numerically in Fig.~\ref{fig:kappa-T}, which exhibits the power-law scalings distinct at low and high temperatures.
Our findings are directly relevant to ultracold atom experiments where the thermal conductivity is measurable~\cite{Baird:2019,Patel:2020,Li:2022,Wang:2022}, and may be useful for better understanding of their nonequilibrium dynamics.
Furthermore, the bulk viscosity of a one-dimensional Bose gas can be studied along the lines of analyses developed recently in Refs.~\cite{Fujii:2018,Nishida:2019,Enss:2019,Hofmann:2020,Fujii:2020,Maki:2020} for Fermi gases in higher dimensions, which will be reported elsewhere~\cite{Tanaka:2022,Nishida:preprint}.

\acknowledgments
The authors thank Luca V.~Delacr\'etaz, Masaru Hongo, and Kazumitsu Sakai for valuable discussions.
This work was supported by JSPS KAKENHI Grants No.\ JP18H05405 and No.\ JP21K03384.

\begin{figure}[t]
\includegraphics[width=0.8\columnwidth]{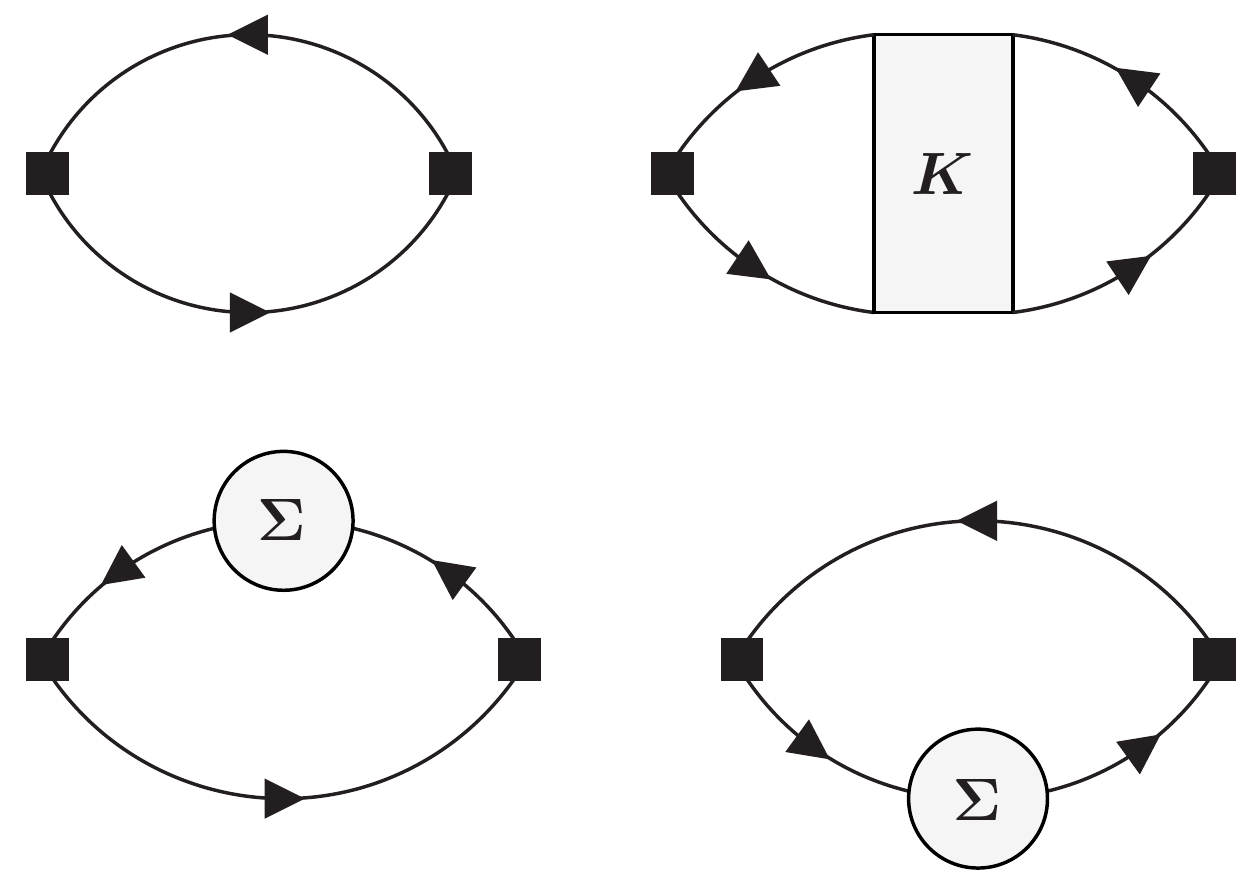}
\caption{\label{fig:perturbation}
Contributions to the imaginary-time-ordered correlation function up to $O(g^2)$ in Eqs.~(\ref{eq:chi_one-loop})--(\ref{eq:chi_perturbation}).
The circle ($\Sigma$) consists of the boson self-energies in Fig.~\ref{fig:self-energy} and the rectangle ($K$) consists of the four-point functions in Fig.~\ref{fig:kernel}.}
\end{figure}

\appendix
\section{Frequency-dependent thermal conductivity and approximate resummation}\label{app:resummation}
As complementary information to Sec.~\ref{sec:microscopic}, we present the frequency-dependent thermal conductivity resulting from Eqs.~(\ref{eq:time-ordered}) and (\ref{eq:self-consistent}) up to $O(g^2)$ and the approximate resummation scheme.
The corresponding imaginary-time-ordered correlation function to the lowest order in perturbation is provided by the one-loop diagram depicted in Fig.~\ref{fig:perturbation},
\begin{align}\label{eq:chi_one-loop}
\chi_Q(iw)|_{O(g^0)} = \frac1\beta\sum_v\int_p\Q_pG(iv+iw,p)G(iv,p)\Q_p,
\end{align}
whereas those at $O(g^2)$ are provided by the self-energy correction,
\begin{align}
& \chi_Q(iw)|_{\Sigma_{22,33}} = \frac1\beta\sum_v\int_pG(iv+iw,p)\Q_p \notag\\
&\quad \times G(iv,p)\Sigma_{22,33}(iv,p)G(iv,p)\Q_p + (iw\to-iw),
\end{align}
and the Maki-Thompson and Aslamazov-Larkin corrections,
\begin{align}\label{eq:chi_perturbation}
& \chi_Q(iw)|_{K^{\MT,\AL}_{22,33}}
= \frac1{\beta^2}\sum_{v,v'}\int_{p,p'}\Q_pG(iv+iw,p)G(iv,p) \notag\\
&\quad \times K^{\MT,\AL}_{22,33}(iv+iw,iv;p|iv'+iw,iv';p') \notag\\
&\quad \times G(iv'+iw,p')G(iv',p')\Q_{p'}.
\end{align}
We will find below that the frequency-dependent thermal conductivity has the structure of $\kappa(\omega)\sim g^0/\omega+g^2/\omega^2+\cdots$ at low frequency, where the corrections to the chemical potential and the contributions from the two-body and three-body operators in Eq.~(\ref{eq:energy-flux}) appear only at subleading orders so as to be suppressed here.

The Matsubara frequency summations are readily performed by replacing them with the complex contour integrations and the resulting expressions can be arranged into
\begin{align}
\chi_Q(iw)|_{O(g^0)}
= -\int_p\frac{f_p-f_{p+k}}{iw+\ep-\eps_{p+k}}\Q_p^2\,\bigg|_{k\to0},
\end{align}
as well as
\begin{widetext}
\begin{align}
\chi_Q(iw)|_{O(g_2^2)} &\equiv \chi_Q(iw)|_{\Sigma_{22}}
+ \chi_Q(iw)|_{K^\MT_{22}} + \chi_Q(iw)|_{K^\AL_{22}} \notag\\
&= -4\g_2^2\int_{p,p',q,q'}\frac{2\pi\delta(p+p'-q-q')}
{(\ep+\eps_{p'}-\eq-\eps_{q'})
\left[(\ep+\eps_{p'}-\eq-\eps_{q'})^2-(iw)^2\right]} \notag\\
&\quad \times \left[f_pf_{p'}(1+f_q)(1+f_{q'}) - (1+f_p)(1+f_{p'})f_qf_{q'}\right]
\Q_p(\Q_p+\Q_{p'}-\Q_q-\Q_{q'}),
\end{align}
\begin{align}
\chi_Q(iw)|_{O(g_3^2)} &\equiv \chi_Q(iw)|_{\Sigma_{33}}
+ \chi_Q(iw)|_{K^\MT_{33}} + \chi_Q(iw)|_{K^\AL_{33}} \notag\\
&= -6g_3^2\int_{p,p',p'',q,q',q''}\frac{2\pi\delta(p+p'+p''-q-q'-q'')}
{(\ep+\eps_{p'}+\eps_{p''}-\eq-\eps_{q'}-\eps_{q''})
\left[(\ep+\eps_{p'}+\eps_{p''}-\eq-\eps_{q'}-\eps_{q''})^2-(iw)^2\right]} \notag\\
&\quad \times \left[f_pf_{p'}f_{p''}(1+f_q)(1+f_{q'})(1+f_{q''})
- (1+f_p)(1+f_{p'})(1+f_{p''})f_qf_{q'}f_{q''}\right] \notag\\
&\quad \times \Q_p(\Q_p+\Q_{p'}+\Q_{p''}-\Q_q-\Q_{q'}-\Q_{q''}),
\end{align}
where terms involving $\Q_p$, $\Q_{p',p''}$, and $\Q_{q,q',q''}$ in parentheses are provided by the self-energy, Maki-Thompson, and Aslamazov-Larkin corrections, respectively.
According to the Kubo formula in Eq.~(\ref{eq:kubo}), each of them contributes to the frequency-dependent thermal conductivity as
\begin{align}\label{eq:kappa_one-loop}
\kappa(\omega)|_{O(g^0)}
= \frac{\Im[\chi_Q(\omega+i0^+)|_{O(g^0)}]}{T\omega}
= \beta^2\int_pf_p(1+f_p)\Q_p^2 \times \pi\delta(\omega),
\end{align}
and similarly,
\begin{align}
\kappa(\omega)|_{O(g_2^2)}
&= -\frac{2\g_2^2}{T\omega^3}\int_{p,p',q,q'}2\pi\delta(p+p'-q-q')\,
\pi[\delta(\ep+\eps_{p'}-\eq-\eps_{q'}-\omega)
- \delta(\ep+\eps_{p'}-\eq-\eps_{q'}+\omega)] \notag\\
&\quad \times \left[f_pf_{p'}(1+f_q)(1+f_{q'}) - (1+f_p)(1+f_{p'})f_qf_{q'}\right]
\Q_p(\Q_p+\Q_{p'}-\Q_q-\Q_{q'}),
\end{align}
\begin{align}
\kappa(\omega)|_{O(g_3^2)}
&= -\frac{3g_3^2}{T\omega^3}\int_{p,p',p'',q,q',q''}2\pi\delta(p+p'+p''-q-q'-q'') \notag\\
&\quad \times \pi[\delta(\ep+\eps_{p'}+\eps_{p''}-\eq-\eps_{q'}-\eps_{q''}-\omega)
- \delta(\ep+\eps_{p'}+\eps_{p''}-\eq-\eps_{q'}-\eps_{q''}+\omega)] \notag\\
&\quad \times \left[f_pf_{p'}f_{p''}(1+f_q)(1+f_{q'})(1+f_{q''})
- (1+f_p)(1+f_{p'})(1+f_{p''})f_qf_{q'}f_{q''}\right] \notag\\
&\quad \times \Q_p(\Q_p+\Q_{p'}+\Q_{p''}-\Q_q-\Q_{q'}-\Q_{q''}).
\end{align}
Although both two-body and three-body contributions are naively $O(\omega^{-2})$ at low frequency, we find that
\begin{align}
\kappa(\omega)|_{O(g_2^2)}
= \frac{\beta^2}{\omega^2}\int_{p,p',q,q'}W_2(p,p'|q,q')
f_pf_{p'}(1+f_q)(1+f_{q'})\Q_p(\Q_p+\Q_{p'}-\Q_q-\Q_{q'}) + O(\omega^0)
\end{align}
cancels out due to the energy and momentum conservations in $W_2(p,p'|q,q')$ but
\begin{align}\label{eq:kappa_perturbation}
\kappa(\omega)|_{O(g_3^2)}
&= \frac{\beta^2}{\omega^2}\int_{p,p',p'',q,q',q''}W_3(p,p',p''|q,q',q'')
f_pf_{p'}f_{p''}(1+f_q)(1+f_{q'})(1+f_{q''}) \notag\\
&\quad \times \Q_p(\Q_p+\Q_{p'}+\Q_{p''}-\Q_q-\Q_{q'}-\Q_{q''}) + O(\omega^0)
\end{align}
remains at the leading order.
\end{widetext}

Because $\kappa(\omega)|_{O(g_3^2)}$ is formally divergent at zero frequency, obtaining the thermal conductivity in Eq.~(\ref{eq:kubo}) requires the resummation.
To this end, we express Eq.~(\ref{eq:kappa_one-loop}) by $\Im[-D/(\omega+i0^+)]$ and regard $\Im[iD/(\omega^2\tau)]$ from Eq.~(\ref{eq:kappa_perturbation}) as a leading correction in a simple geometric series~\cite{Gotze:1972,Enss:2011,Nishida:2019,Hofmann:2020}.
Consequently, the frequency-dependent thermal conductivity is brought into the Drude form of
\begin{align}\label{eq:drude}
\kappa(\omega) \approx \Im\!\left(\frac{-D}{\omega+i/\tau}\right)
= \frac{D\tau}{(\omega\tau)^2+1},
\end{align}
where $D$ is the Drude weight in the noninteracting limit and $\tau$ is the relaxation time introduced in Eq.~(\ref{eq:relaxation}).
Because the thermal conductivity is found to be $\lim_{\omega\to0}\kappa(\omega)=D\tau$, the approximate resummation scheme adopted herein is equivalent to solving the linearized Boltzmann equation with the relaxation-time approximation as described in Sec.~\ref{sec:thermal}.
In particular, the thermal conductivity diverges as $\kappa(\omega)=D\pi\delta(\omega)$ in the limit of vanishing three-body interaction, $\tau\to\infty$, because the two-body interaction does not contribute to the finite relaxation time in one dimension.

Finally, we note that the frequency-dependent thermal conductivity of nonintegrable but momentum-conserving systems is considered to be divergent as $\kappa(\omega)\sim\omega^{-1/3}$ at sufficiently low frequency $\omega\ll1/\tau_\hd$ due to enhanced hydrodynamic fluctuations in one dimension~\cite{Narayan:2002,Dhar:2008,Samanta:2019}.
If this is the case, our finite thermal conductivity $\sim D\tau$ is valid only at $1/\tau_\hd\ll\omega\ll1/\tau\sim O(g_3^2)$ assuming the existence of such a frequency window.
Because all leading-order divergences up to $O(g^2)$ are resumed into Eq.~(\ref{eq:drude}), hydrodynamic fluctuations should be subleading or higher-order effects in perturbation.
Whether and how such divergence resulting from hydrodynamic fluctuations arises in our microscopic approach remains an important problem to be resolved in future work.

\end{document}